\documentclass[aps,prc,twocolumn,groupedaddress,showpacs]{revtex4}

\usepackage{graphicx}
\usepackage[rightcaption]{sidecap}

\begin{document}
\title{ Unification of Airy structure  in inelastic
$\alpha$ +$^{16}$O scattering \\ and  $\alpha$-cluster structure
 with core excitation  in $^{20}$Ne
 }

\author{
 Y. Hirabayashi$^1$ and  S. Ohkubo$^{2,3}$ 
 }
\affiliation{$^1$Information Initiative Center,
Hokkaido University, Sapporo 060-0811, Japan}
\affiliation{$^2$Research Center for Nuclear Physics, Osaka University, 
Ibaraki, Osaka 567-0047, Japan }
\affiliation{$^3$University of Kochi, 5-15 Eikokuji-cho, Kochi 780-8515, Japan  }

\date{\today}

\begin{abstract}
\par
The Airy structure of the nuclear  rainbow and prerainbow in inelastic  and 
elastic $\alpha$+$^{16}$O  scattering is studied with the coupled channel method
 using a folding potential derived from the microscopic  wave functions 
of $^{16}$O. The theoretical calculations reproduce  the characteristic 
energy evolution of the Airy minimum of the experimental angular  distributions. 
 The energy levels with $\alpha$-cluster structure in $^{20}$Ne are  reproduced well using the
 potentials determined from the analysis of scattering.
It is shown that the emergence of the $K=0_3^+$ $\alpha$-cluster band with core excitation at 7.19 MeV 
in $^{20}$Ne   is intimately related to the emergence of the prerainbow
 and rainbow in inelastic scattering to the $^{16}$O($0^+_2$). 
It is found that the   $\alpha$-cluster states with core excitation, 
 the  prerainbow and the rainbow in inelastic scattering are understood in a unified way
as well as in  the case of elastic scattering.  
\end{abstract}


\maketitle

\section{INTRODUCTION}
\par  
Alpha-cluster  structure exists widely  and essential for understanding
nuclear properties  in light and medium-weight nuclei \cite{Suppl1972,Suppl1980,Ohkubo1998,Michel1998}.
The typical  $^{20}$Ne nucleus  with two  protons and two neutrons on top of the inert double 
magic closed shell nucleus $^{16}$O  has an  $\alpha$+$^{16}$O cluster structure  
 and has been studied thoroughly with a  cluster model  \cite{Suppl1972,Suppl1980,Ohkubo1998,Michel1998}.
In understanding the $\alpha$-cluster structure of nuclei the  interaction potential between
 the  $\alpha$ particle and the nucleus is very important \cite{Michel1998}.
The nuclear rainbow  can be observed when the absorption is weak or incomplete and 
the systematic study of nuclear rainbow scattering makes it possible to determine
 the interaction potential up to the  internal region \cite{Khoa2007}.

 \par
The  elastic $\alpha$ particle scattering
  from $^{16}$O has been studied over a wide range of incident energies 
theoretically and experimentally  \cite{Michel1998} and it has been shown that 
the  global optical potential  can  describe  rainbow scattering,
 prerainbows, anomalous large angle scattering (ALAS)  
 in the  low energy region, and the $\alpha$+$^{16}$O cluster structure of $^{20}$Ne 
 in a unified way \cite{Ohkubo1977,Michel1983,Abele1993,Michel1998}. The  characteristic 
 evolution of the
 angular distribution from
 the low energy  region where $\alpha$-cluster structure is relevant to the high energy
 region where the typical nuclear rainbow appears can be  understood very well
 systematically.  The {\it raison d'etre} of the $\alpha$-cluster structure in the ground
 state of $^{20}$Ne is thus found to  be closely related to the existence of the nuclear rainbows
 for the $\alpha$+$^{16}$O system. 
 The   emergence of the $\alpha$+$^{16}$O structure  in the ground state of $^{20}$Ne is  a consequence 
 of the interaction potential which  causes the nuclear rainbow for the $\alpha$+$^{16}$O system.
 This unified understanding was also   successfully applied to 
the  nuclear rainbow  in elastic $\alpha$ particle scattering from $^{40}$Ca
 and the  $\alpha$-cluster  structure of  $^{44}$Ti 
\cite{Michel1986,Michel1998}.

\par  
Compared with elastic rainbow scattering,
inelastic  rainbow scattering has  not been    studied  extensively over a wide
 range of incident energies  both experimentally and theoretically.
However, similar to the  rainbow in elastic scattering, the study of the inelastic
 rainbow scattering will be very useful in understanding the interaction potential
 for the inelastic channels 
\cite{Bohlen1982,Bohlen1985,Bohlen1993,Khoa2005,Khoa2007}.
The mechanism of the nuclear rainbow and   the  Airy structure in inelastic scattering  has been 
 studied  for the $\alpha$+$^{40}$Ca  and $^6$Li+$^{12}$C systems  by using a
 phenomenological  form factor \cite{Michel2004A,Michel2005}.
 On the other hand, from the viewpoint of a nuclear structure study  it has recently 
been shown that 
 inelastic nuclear rainbow scattering is  powerful in understanding  the $\alpha$-cluster structure
 of the excited states of the  nucleus. For example,  the $\alpha$ 
particle  condensate  states in  $^{12}$C and $^{16}$O have been revealed by studying the 
  rainbow  and prerainbow in $\alpha$ particle and $^3$He scattering from $^{12}$C \cite{Ohkubo2004,Ohkubo2007,Ohkubo2010,Hamada2013}.
However, a unified study of inelastic rainbow scattering and  $\alpha$-cluster 
 structure has never been undertaken for the typical $\alpha$+$^{16}$O system.

\par
The purpose of this paper is to understand the nuclear rainbow and prerainbow  in inelastic 
$\alpha$+$^{16}$O scattering,  and the $\alpha$-cluster structure  in $^{20}$Ne  in a unified way.
 It is shown  that the emergence of  $\alpha$-cluster structure with core
 excitation in $^{20}$Ne   is closely related to  the appearance of  Airy structure in the 
 prerainbow and rainbow inelastic $\alpha$ particle scattering from $^{16}$O. 

\begin{SCfigure*}
\includegraphics[keepaspectratio,width=12cm] {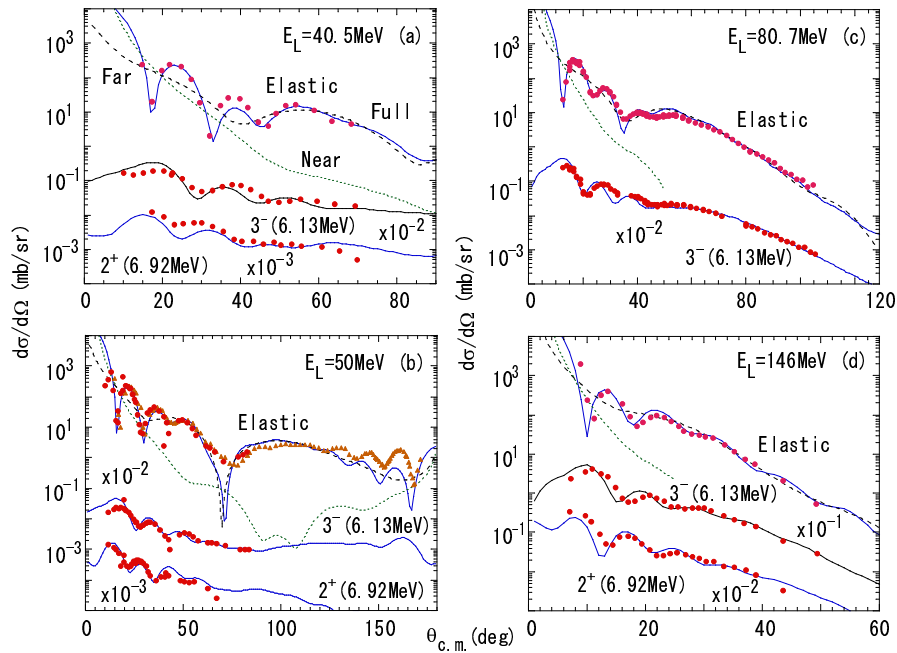}
\protect\caption{\label{fig.1} {(Color online) 
The  elastic and inelastic  angular distributions (solid lines)  of $\alpha$ 
scattering from $^{16}$O calculated   using the coupled channel
 method are  compared with    the experimental   data (red points)  at  $E_L$=40.5 MeV
\cite{Harvey1966},   50 MeV,  80.7 MeV   \cite{Reed1968} and  146 MeV     \cite{Knopfle1975}.
In (b)  the elastic scattering angular distribution  calculated at 48.7 MeV is displayed  to be compared with the
 data (orange triangles) measured up to very large angles   at 48.7 MeV \cite{Abele1993}.
 The calculated cross sections for elastic scattering  (solid lines) are shown 
decomposed into the farside component (dashed lines) and the nearside component (dotted lines).
 }
}
 \end{SCfigure*}

\section{DOUBLE FOLDING MODEL}

We  study the elastic and inelastic  angular 
 distributions of  $\alpha$+$^{16}$O   scattering systematically 
 with the coupled channel method  using a microscopic double folding  model.
In the coupled channel calculations we simultaneously take into account  the 
 0$^+_1$ (0.0 MeV), $0_2^+$ (6.05 MeV), 3$^-$ (6.13 MeV) and  
  2$^+$ (6.92 MeV) states of $^{16}$O.
The double folding (DF) potential is constructed from the transition densities
 of $^{16}$O using a density-dependent effective interaction:
\begin{eqnarray}
\lefteqn{V_{ij}({\bf R}) =
\int \rho_{00}^{\rm (\alpha)} ({\bf r}_{1})\;
     \rho_{ij}^{\rm (^{16}O)} ({\bf r}_{2})} \nonumber\\
&& \times v_{\rm NN} (E,\rho,{\bf r}_{1} + {\bf R} - {\bf r}_{2})\;
{\rm d}{\bf r}_{1} {\rm d}{\bf r}_{2} ,
\end{eqnarray}
\noindent
where $\rho_{00}^{\rm (\alpha)} ({\bf r})$ is the ground state density
of the $\alpha$ particle, while $v_{\rm NN}$ denotes the density dependent
M3Y effective interaction (DDM3Y) \cite{Kobos1984} usually used in the DF model.
$\rho_{ij}^{\rm (^{16}O)} ({\bf r})$ represents the diagonal ($i=j$) or
transition ($i\neq j$) nucleon density of $^{16}$O
which is obtained  from  the microscopic wave functions calculated 
 in the $\alpha$+$^{12}$C cluster model  using the orthogonality
 condition model (OCM) \cite{Okabe1995}.
The OCM wave functions we take here have been configured by
 using a realistic  size for the $\alpha$ particle and  $^{12}$C.
 As a result  the agreement 
 of the theoretical  calculation
 with the experiment  is    further improved from the  original 
 $\alpha$+$^{12}$C cluster model wave functions by Suzuki \cite{Suzuki1976}, which already 
 excellently reproduced  almost all the energy levels of $^{16}$O  up to $E_x$$\approx$13 MeV. 
This cluster model simultaneously reproduces  not only  the $\alpha$-cluster states like 
 the $K=0^+$ band starting from the mysterious $0^+$ state at 6.05 MeV, but also the shell-model-like states
 such as the 
   3$^-$ (6.13 MeV) state and the  ground state. 
 In the calculations we introduce the normalization factor  $N_R$ for 
 the real part of the  DF potential \cite{Satchler1979,Brandan1997,Khoa2001}.  We take into account  the important transition densities 
available in Ref.\cite{Okabe1995}, i.e., g.s $\leftrightarrow$ $0^+_2$ (6.05 MeV), $3^-$ (6.13 MeV), 
$2^+$ (6.92 MeV), and  $0^+_2$ (6.05 MeV) $\leftrightarrow$ $2^+$ (6.92 MeV) in addition to all the
 diagonal couplings.  
 The effect of absorption due to couplings to the other channels is  introduced  as a 
 phenomenological imaginary   potential  with a volume-type Wood-Saxon form factor.
  In  previous papers \cite{Ohkubo2004,Ohkubo2007,Ohkubo2010,Hamada2013}
 this method  was successfully applied to the rainbow 
and prerainbow  scattering of $\alpha$ particles and $^3$He from $^{12}$C, where  
realistic wave functions of $^{12}$C  calculated in the microscopic $\alpha$-cluster model 
were used.

\begin{SCfigure*}
\includegraphics[keepaspectratio,width=10cm] {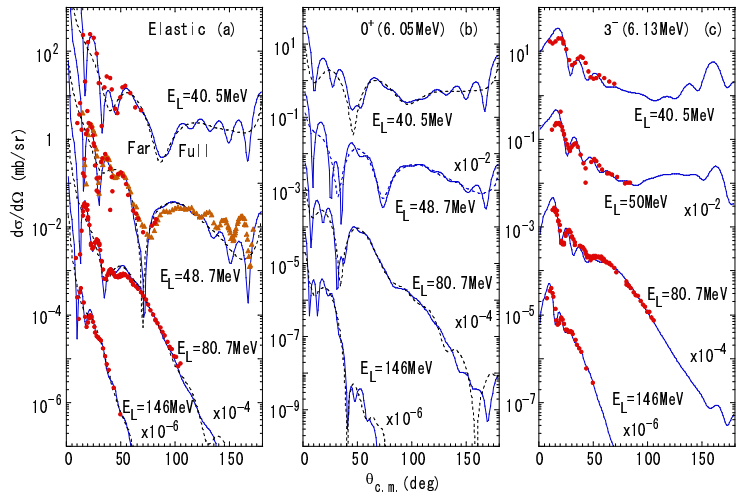}
 \protect\caption{\label{fig.2} {(Color online) 
The energy evolution of  the Airy minimum in the calculated (solid lines) and experimental 
(red points and orange triangles)  angular distributions of prerainbow and   rainbow   $\alpha$+$^{16}$O 
scattering,
is displayed for  (a) elastic scattering,  (b) inelastic scattering to the
  $0_2^+$ state and (c) inelastic scattering to the $3^-$ state.  The  farside cross sections 
    are shown  with dashed lines. The experimental data (red points) at 50 MeV are also included 
for the   48.7 MeV  case in (a).
 }
}
\end{SCfigure*}

\section{AIRY STRUCTURE IN ELASTIC AND INELASTIC $\alpha$+$^{16}$O SCATTERING}
 
We analyze the angular distributions of elastic and inelastic $\alpha$ particle
 scattering from $^{16}$O. Although there are many experimental data available for 
 elastic scattering, experimental angular distributions  for inelastic prerainbow and rainbow 
scattering are only available at the limited energies, 
40.5 MeV \cite{Harvey1966}, 50 MeV,  80.7 MeV \cite{Reed1968} and 146 MeV \cite{Knopfle1975}. 
In Fig.~1 the calculated angular distributions  are  displayed 
  in comparison with the experimental data. 
The agreement with experimental data, which shows a characteristic energy evolution,
  is very good. Here the normalization 
factor for the real part of the potential is slightly adjusted  to take  account of the
 energy dependence of the $N_R$ \cite{Satchler1979,Brandan1997,Khoa2001}. 
The real potential which reproduces the Airy minimum of rainbow scattering  at the highest 
 incident energy 146 MeV is uniquely determined without discrete ambiguity and the obtained 
 volume integral  per nucleon pair for elastic channel is   304 MeVfm$^3$.
 The   $N_R$ values  at the lower energies are determined by 
slightly adjusting to fit  the experimental data.
The imaginary
 potential is mostly responsible for reducing  the magnitude of the cross sections. 
In the calculations the strength parameter of the imaginary potential ($W_V$) was fitted
 to  reproduce  the magnitude of the experimental cross sections while the radius parameter 
 and the diffuseness parameter  were fixed at around $R_V$=5 fm and $a_V$=0.3-0.5 fm,
 respectively. These approaches using the folding potential have been  shown successful
 for the $\alpha$ and $^3$He  scattering  from  $^{12}$C \cite{Ohkubo2004,Ohkubo2007,Hamada2013}  
over a wide rage of incident energies.
The properties of the real folding potential and potential parameters used in the 
 analysis are given in Table I.
The energy dependence of   the volume integrals for the
  elastic channel  is physically reasonable  and    consistent with the phenomenological potentials \cite{Michel1983} 
and the folding model  potentials  
 \cite{Abele1993,Khoa2001} obtained in the analysis of elastic  $\alpha$ particle scattering from $^{16}$O. 
 For elastic scattering the calculated angular distributions are decomposed into farside
 and nearside components. The angular distributions are dominated by the  refractive farside
 component and the Airy  minimum  
is observed at $\sim$18$^\circ$,  $\sim$36$^\circ$ and  $\sim$78$^\circ$ for $E_L$=146 MeV, 80.7 MeV and
 48.7 MeV, respectively.  For 40.5 MeV,  although the experimental data are available only up to 
$\sim$68$^\circ$, the theoretical calculation predicts an Airy minimum at $\sim$85$^\circ$.

\begin{table}[bh]
\begin{center}
\caption{The  volume integral per nucleon
pair $J_V$, normalization factor $N_R$, root mean square radius $<R^2>^{1/2}$ of the
folding potential,  and  the  parameters of the imaginary potentials in the conventional
notation.
 }
\begin{tabular}{cclccccc}
 \hline
  \hline
  $E_L$ & $N_R$ & $J^\pi$ &$J_V$ &  $<R^2>^{1/2}$ & $W_V$ & $R_V$ & $a_V$ \\
  (MeV) & & &      (MeV fm$^3$) & (fm)&(MeV) & (fm) & (fm)\\
 \hline
 40.5 & 1.42 & $0_1^+$  &  395 &  3.61 & 5.0  & 5.1 & 0.5 \\
      &      & $0_2^+$  & 425 &  3.77 & 5.0  & 5.1 & 0.5 \\
      &      & $3^-$    & 394 &  3.58 & 5.0  & 5.1 & 0.5 \\
      &      & $2^+$    & 429 &  3.79 & 8.0 & 5.1 & 0.5 \\
  50 & 1.42 & $0_1^+$  &  390 &  3.61 & 5.5  & 5.1 & 0.3 \\
      &      & $0_2^+$  & 418 &  3.77 & 6.0  & 5.1 & 0.3 \\
      &      & $3^-$    & 388 &  3.58 & 6.0  & 5.1 & 0.3 \\
      &      & $2^+$    & 422 &  3.79 & 11.0 & 5.1 & 0.3 \\
  80.7& 1.34 & $0_1^+$   &  347 &  3.62 & 7.1  & 5.2 & 0.4 \\
      &      & $0_2^+$  &  373 &  3.78 & 10.0 & 5.2 & 0.4 \\
      &      & $3^-$    &  346 &  3.59 & 10.0 & 5.2 & 0.4 \\
      &      & $2^+$    &  377 &  3.79 & 10.0 & 5.2 & 0.4 \\
  146 & 1.34 & $0_1^+$  &  304 &  3.65 & 9.0  & 5.2 & 0.4 \\
      &      & $0_2^+$ &  328 &  3.80 & 13.0 & 5.2 & 0.4 \\
      &      & $3^-$   &  303 &  3.62 & 13.0 & 5.2 & 0.4 \\
      &      & $2^+$   &  331 &  3.82 & 12.0 & 5.2 & 0.4 \\
 \hline
 \hline
\end{tabular}
\end{center}
\end{table}

In Fig.~2 the energy evolution of the Airy structure in the angular distributions for 
elastic scattering is shown  in comparison with that of inelastic scattering
 to the $0_2^+$ state  and the $3^-$ state. 
 As seen in Fig.~2(a) the Airy minimum moves toward large angles as the incident energy
decreases  from rainbow scattering to prerainbow scattering. This is reasonable considering that
 the refractive index increases and  refraction becomes stronger as the incident energy decreases. 
 
The prerainbow at 40.5 MeV and around 50 MeV  develops into the typical rainbow at the higher energies, 80.7 MeV 
 and 146 MeV  with the falloff    
of the cross sections in the angular distribution at the darkside beyond the rainbow angle  and 
with the first order  Airy minimum and Airy maximum of  the bright side at  angles smaller than the rainbow angle.
In Fig.~2(c)  the  nuclear rainbow  is  observed to behave similarly in inelastic 
scattering to the 3$^-$ (6.13 MeV) state at 80.7 and 146 MeV, which shows that the absorption  is not
 strong for inelastic scattering.  This suggests that the inelastic rainbow scattering
 can serve to  determine  the interaction potential for inelastic scattering
 and the transition form factors including the internal region. 
The very good agreement of the theoretical calculations with the 
experimental data  shows that the present potential for the inelastic channel derived
 from the OCM microscopic  cluster wave functions is reliable  up to the internal region.
The Airy minimum for the rainbow  scattering to the 3$^-$ state is observed at
 $\sim$27$^\circ$ and $\sim$44$^\circ$ for 146 MeV  and 80.7 MeV, respectively.
The  fall-off of the angular distribution characteristic to the 
rainbow at the high energy region is seen at 146 MeV  and 80.7 MeV for the inelastic scattering
 to the  $3^-$ state  and  the Airy minimum of the indication of the prerainbow is seen
 at around 88$^\circ$ in the 50 MeV angular distribution, although the experimental data
are  available only up to 83$^\circ$.

\par
The  good agreement  of the calculated angular distributions  with the experimental 
data in Fig.~1 shows that the interaction potentials  constructed from the OCM wave function  are  reliable up 
to the internal  region. This makes it  possible to investigate the energy evolution of the Airy minimum
 for the $0_2^+$ (6.05 MeV) state reliably. Although the experimental data have not been measured, we see in
Fig.~2(b) that the evolution of the angular distributions for this state is
 very similar to  that for  elastic scattering. 
It is interesting to note that although the two 0$^+$ states have a very different
 structure, shell-like spherical for the ground state and the deformed well-developed
 $\alpha$-cluster structure for the $0^+_2$ state, the  essential behavior of the two
 angular distributions is similar.  With refractive scattering the target nucleus behaves 
 as a lens. The similarity may be related to the fact that the  difference in the sizes
 of the two states is not large: the calculated  rms radius of the matter density distribution 
 is 2.58 fm for the ground state and 2.77 fm for the $0_2^+$ state. 
 
\begin{figure}[t]
\includegraphics[keepaspectratio,width=8.6cm] {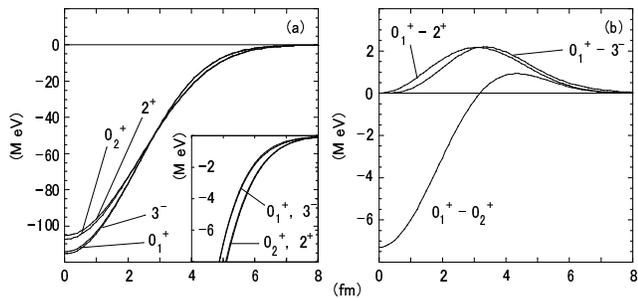}
 \protect\caption{\label{fig.3} { 
The double folding diagonal (a) and coupling  (b)  potentials for the  $\alpha$+$^{16}$O system calculated
at $E_L$=50 MeV}.
}
\end{figure}
 \par
In Fig.~3  the diagonal and coupling interaction potentials are displayed.  
The diagonal potential for the $3^-$ state is similar to that for the ground state in 
 magnitude and  shape.  We see  in Table I that  the volume integrals and rms radii of
 the potentials  for the ground state  and the $3^-$ state are very similar.  These two
 states have a  compact shell-model structure. On the other hand, the diagonal  potentials
 for the
$0^+_2$ and $2^+$ states are significantly shallower in the internal
region and  deeper in the surface region compared with those for the ground  and 
$3^-$ states.
  This  is due to the fact  that the $0^+_2$ and $2^+$ states have a well-developed 
 $\alpha$+$^{12}$C  cluster structure. 
However, this difference of the interaction potential is important when
 we  want to understand the  bound and quasi-bound states   of the    $\alpha$+$^{16}$O system,
  the  low energy prerainbow scattering and  the high-energy  rainbow scattering
 in an inelastic channel in a unified way,
 although in  phenomenological studies the same potential  is often used for elastic
  and inelastic channels. It is also important to point out that the form factor
 of the coupling from the ground state to the $0^+_2$ state  has a node and is different 
from a phenomenological monopole  vibrational form factor derived from the Woods-Saxon potential.

\begin{figure}[tbh]
 \includegraphics[keepaspectratio,width=8.6cm] {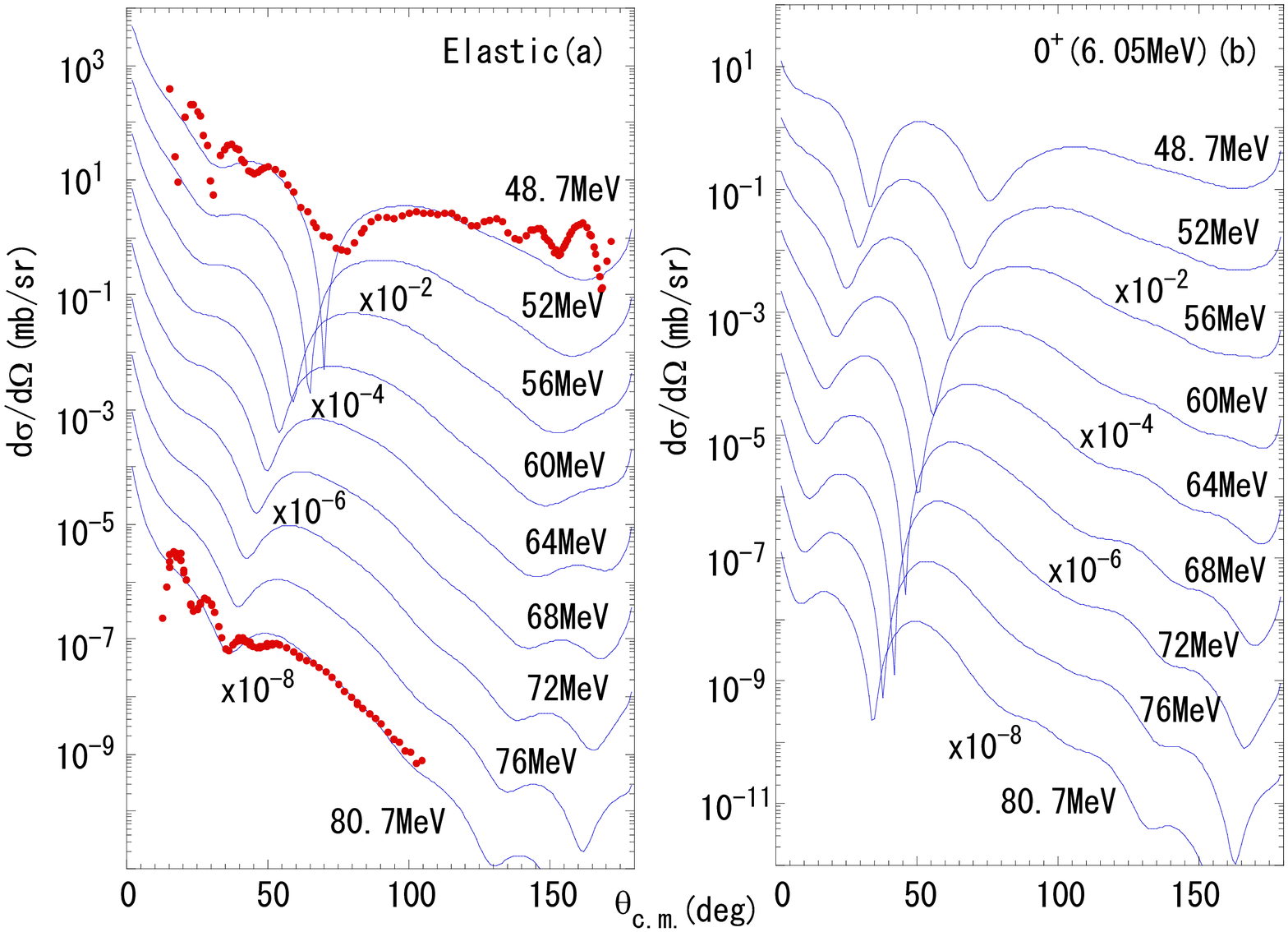}
\protect\caption{\label{fig.4} {(Color online) 
The energy evolution of the  farside component of the  angular distributions calculated
 in 4 MeV steps for (a) elastic scattering 
 and (b) inelastic scattering  to the $0^+_2$ state  (solid lines) and the  experimental data
(points) \cite{Abele1993,Reed1968}.
 }
}
\end{figure}

In Fig.~4 the energy evolution of the farside component of the calculated angular 
 distributions for elastic scattering and inelastic scattering to the $0^+_2$ state is 
compared in 4 MeV steps.
The energy dependence of $N_R$ and  parameters of  the imaginary potentials in-between 
is interpolated from those which fit the
 experimental angular  distributions at 48.7 MeV and 80.7 MeV.  The evolution of the
 Airy minimum, which shifts  to  forward angles as the incident energy increases, is clearly
 seen.  In the inelastic scattering the second order Airy minimum is  seen. 
The similarity of the evolution between the  elastic scattering and the  inelastic
 scattering to the $0^+_2$ state persists from  the lowest  energy to the highest energy.
 This similarity  also persists to the  lower energy ALAS region.
 Although  no experimental data for the $0^+_2$ state  are available, angular distributions 
similar to the elastic scattering, if measured, are expected.  The observation will be very
  useful for  clarifying the coupling form factor  between the two $0^+$ states 
with a very different structure experimentally.

\begin{SCfigure*}
\includegraphics[keepaspectratio,width=10cm] {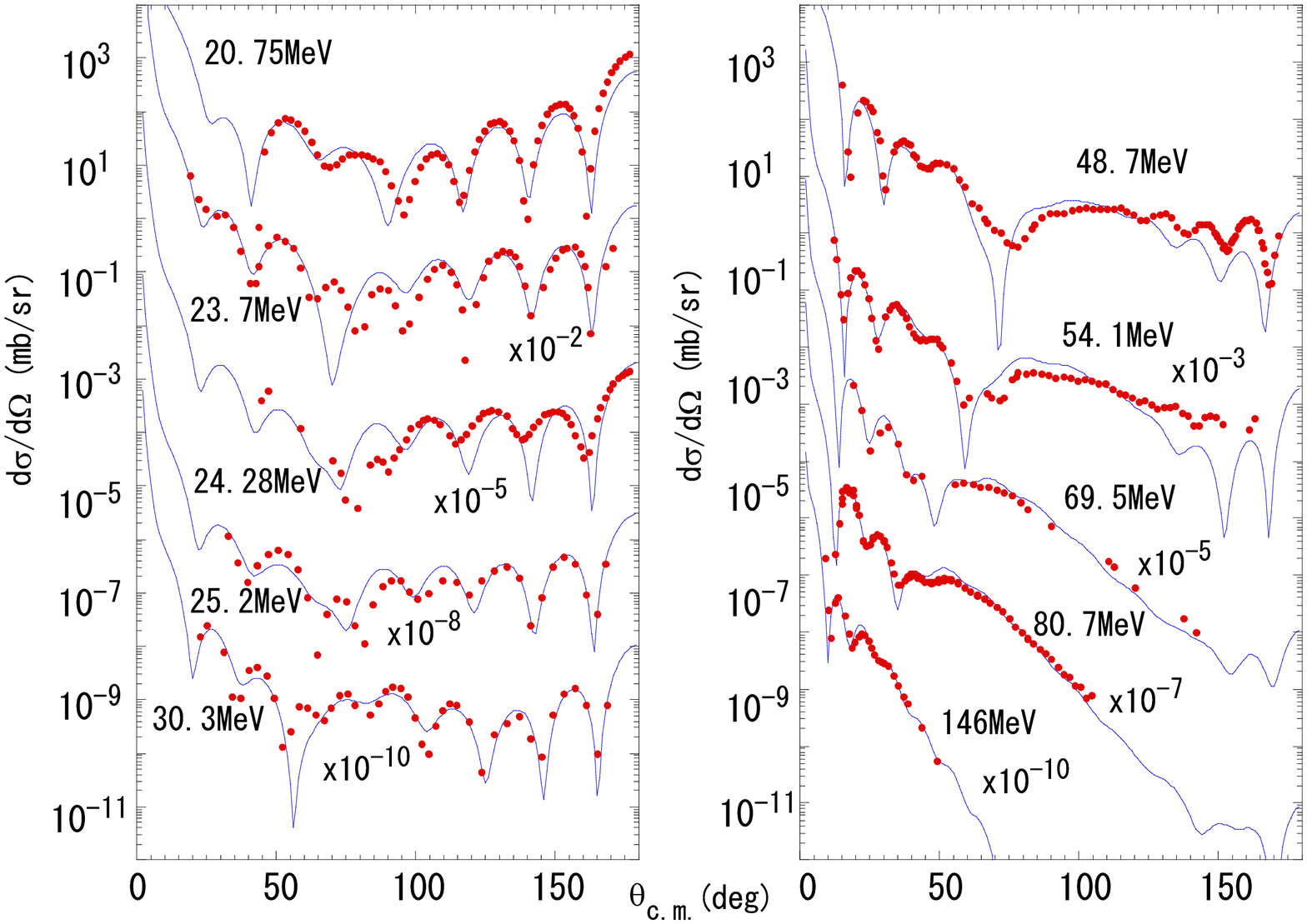}
 \protect\caption{\label{fig.5} {(Color online) 
The energy evolution of  the angular  distributions  in elastic
$\alpha$ + $^{16}$O   scattering calculated  using the coupled channel
 method 
(solid lines) is displayed in comparison with  the experimental data  (points),
 20.7 MeV \cite{Takeda1971}, 23.7 MeV \cite{Feofilov1976}, 
24.28 MeV \cite{Takeda1971}, 25.2 MeV \cite{Ignatenko1996}, 30.3 MeV 
\cite{Ignatenko1996}, 48.7 MeV, 54.1 MeV  \cite{Abele1993}, 69.5 MeV \cite{Michel1983},
50 MeV, 80.7 MeV \cite{Reed1968} and 146 MeV \cite{Knopfle1975}.
 }
}
\end{SCfigure*}

\section{ ALPHA-CLUSTER STRUCTURE IN $^{20}$N$\lowercase{e}$}

\par
In Fig.~5  we show that the present folding potentials using the microscopic
  wave functions can reproduce the elastic scattering up to the  low energy
 region where ALAS appears systematically.  The imaginary potential  simply
takes into account the reduction of the flux due to absorption and the essential structure
 of the angular  distributions is characterized by the real part of the optical 
potential. The $N_R$ values used are reasonably consistent with Table I and  are
1.42, 1.38, 1.42, 1.42, 1.36, 1.32, 1.36 and 1.42 for  
  69.5 MeV, 54.1 MeV, 48.7 MeV, 30.3 MeV, 25.2 MeV, 24.28 MeV, 23.7 MeV and 20.75 MeV, respectively.
 The pronounced oscillation of the angular distributions at the backward
 angles at  20$\sim$25 MeV is due  to the highly excited $\alpha$+$^{16}$O cluster structure
 in $^{20}$Ne  \cite{Ohkubo1977,Michel1983}. 
This suggests the present potential is useful even for the much lower energy region including
 the bound energy region. In fact, the lowest Pauli allowed  states  that  the real potential 
for the elastic channel  accommodate satisfy the Wildermuth condition $2n+L=8$  
where $n$ is the number of the nodes and $L$ is the orbital angular momentum of the relative
 wave function  and correspond to 
the   $K=0_1^+$ band  of $^{20}$Ne (Fig.~6) with the  $\alpha$+$^{16}$O  cluster 
structure.

\begin{figure}[b]
\includegraphics[keepaspectratio,width=8.6cm] {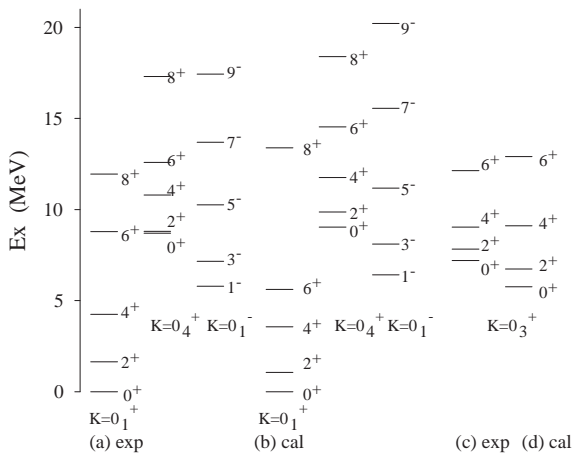}
 \protect\caption{\label{fig.6} {
 The experimental  $K=0_1^+$,  $K=0_1^-$ and $K=0_4^+$ bands with the
 $\alpha$+$^{16}$O(g.s.)  structure (a), and the $K=0_3^+$ band with the 
$\alpha$+$^{16}$O$^*$($0_2^+$) structure (c) are compared with the calculated energy
 levels with the folding potentials for the  $\alpha$+$^{16}$O (b) and 
 $\alpha$+$^{16}$O$^*$($0_2^+$) (d) channels, respectively.
 }
}
\end{figure}

\par
 In  Fig.~6(b) and Fig.~6(d) the energy levels of $^{20}$Ne  calculated in the bound state
 approximation    using the diagonal potentials with $N_R$=1.245  for the elastic and 
the $0_2^+$ channels, respectively,  are shown in comparison with the experimental levels.
 The  $N_R$ used  is the one adjusted to reproduce the experimental binding energy of the ground
 state of $^{20}$Ne, 4.73 MeV,  from the $\alpha$ threshold. 
The calculation  reproduces
 the experimental  energy levels of the $K=0_1^+$   ground band, its parity doublet partner $K=0^-$ 
band  and the  higher nodal $K=0_4^+$  band  with the $\alpha$ +$^{16}$O(g.s.) 
cluster structure well. 
In Fig.~6(d) we see that the  calculated lowest $0^+$ state with the $\alpha$+$^{16}$O($0^+_2$)
 structure     corresponds well to the experimental  $K=0_3^+$   band  starting at $E_x$=7.19 MeV 
(Fig.6(c)) with the $^{12}$C+$\alpha$+$\alpha$ cluster \cite{Suppl1980}.
 The agreement with the experimental $K=0_3^+$ band will be improved by taking into account the coupling
between the two channels because the calculated  $0^+$ state with the $\alpha$+$^{16}$O($0^+_2$) structure 
is pushed
 higher due to the orthogonality to the ground state.
The excitation energy 7.19 MeV of the $0_3^+$   in $^{20}$Ne is close to the excitation energy 6.05 MeV
 of the $0_2^+$ state in $^{16}$O.  In this unified description of the  $K=0_3^+$  band and inelastic scattering
it is very  important that the potential for the $0^+_2$ channel is slightly  shallower in the internal region
 compared with the elastic channel, as shown in Fig.~3.
The interaction potential which describes well the inelastic rainbow scattering for the $\alpha$+$^{16}$O
 inevitably predicts the existence of an $\alpha$-cluster structure with core excitation  near 
 the threshold energy supporting the Ikeda's threshold rule \cite{Suppl1980} even for a core-excited cluster 
case.
 Thus the emergence of the $\alpha$-cluster structure  with core excitation in the bound state energy
 region is  considered to be  in 
line  with  the rainbow, prerainbow  and ALAS for the inelastic 
channel as in the case  for the  elastic channel.

\section{SUMMARY AND CONCLUSIONS}

\par
 To summarize, we  analyzed the nuclear rainbow, prerainbow and ALAS 
in inelastic scattering of $\alpha$ particles from $^{16}$O as well as elastic scattering 
in the  coupled channel method by using a double folding potential derived from 
 microscopic cluster wave functions. The calculations reproduce the experimental angular distributions
 well over a wide range of incident energies and can explain the energy evolution of the
 Airy  minimum in the ALAS, prerainbow and rainbow systematically. The theoretical 
calculations  predict  a clear  nuclear rainbow and  prerainbow  
 in inelastic $\alpha$ particle scattering to the $0^+_2$ (6.05 MeV) state of
 $^{16}$O that resembles  the elastic scattering.
 The interaction potential for the inelastic channel can be well determined
 from the analysis of inelastic nuclear rainbow scattering as was the case for the elastic
 scattering.  This indicates that the interaction potential for the  inelastic channel
  can also describe the $\alpha$-cluster state of $^{20}$Ne with the $^{12}$C+$\alpha$ core excitation.
  Our  potential  locates the $K=0_3^+$ $\alpha$-cluster band with core excitation
 in $^{20}$Ne  in good agreement with experiment in addition to the  $K=0_1^+$,  $K=0_1^-$ 
  and  $K=0_4^+$ bands with the   $\alpha$+$^{16}$O(g.s.) structure.
  In conclusion, we have shown for the first time 
that the $\alpha$-cluster structure with core excitation, the ALAS, the prerainbow and the nuclear rainbow
with its beautiful energy evolution of the Airy structure can be understood in a unified way.
 Inelastic nuclear rainbow scattering is useful not only for extracting the
 interaction potential but also for the understanding of the $\alpha$-cluster structure
 with core excitation in the bound state region.
  
\section{ACKNOWLEDGMENTS}

The authors thank S.~Okabe for providing us with the transition densities.
 One of the authors (SO) thanks the Yukawa Institute for Theoretical Physics for
 the hospitality extended  during a stay in February 2013. 
Part of this work was 
supported by the Grant-in-Aid for the Global COE Program ``The Next
 Generation of Physics, Spun from Universality and Emergence'' from the Ministry 
of Education, Culture, Sports, Science and Technology (MEXT) of Japan.

\end{document}